\documentclass[12pt]{article}

\usepackage{amsmath,graphicx,amsthm}

\newcommand{\f}{\operatorname}

\title{Maximum Likelihood Estimation for the Weight Lindley Distribution Parameters under Different Types of Censoring}
\author{P. L. RAMOS,\thanks{ Email: pedrolramos@usp.br}  
\ \ \ \ F. LOUZADA\thanks{ Email: louzada@icmc.usp.br} \ \  \ \ V. G. CANCHO\thanks{ Email: garibay@icmc.usp.br} \\ \\ Institute of Mathematical Science and Computing \\ Universidade de Sao Paulo, Sao Carlos-SP, Brazil } 
\date{\today}

\begin{document}
\maketitle

\begin{abstract}
In this paper the maximum likelihood equations for the parameters of the Weight Lindley distribution are studied considering different types of censoring, such as, type I, type II and random censoring mechanism. A  numerical simulation study is perform to evaluate the maximum likelihood estimates.  The proposed methodology is illustrated in a real data set.

\end{abstract}

\section{Introduction}

Advances in computational methods and numerical simulations have allowed to incorporate efficient models that are capable of describing real problems. Introduced by Ghitany et. al (2011) the Weight Lindley distribution with two parameters, is very flexible model to be fitted by reliability data since this distribution has increasing and bathtub hazard shape.

Some properties of this model were studied by Ghitany et. al (2011) as well as the parameter estimation based on the maximum likelihood method. Mazucheli et al. (2013) compare the efficiency of four estimation methods: maximum likelihood, method of moments, ordinary least-squares, and weighted least-squares and conclude that the weighted least-squares method reproduces similar results to those obtained using the maximum likelihood. Using a Bayesian approach Ali (2013) consider different non-informative and informative prior for the parameters of the WL distribution.

However, in studies involving a temporal response, is common the presence of incomplete or partial data, the so called censored data (Lawless, 2002). It is important to point out that even incomplete these data provide important information about the lifetime of the components and the omission of those can result in biased conclusions. In literature there are different mechanisms of censorship (Balakrishnan \& Aggarwala, 2000; Lawless, 2002; Balakrishnan \& Kundu, 2013). Due to the large number of applications in medical survival analysis and industrial life testing, it will be considered censored data with type II, type I and random censoring mechanism. Some referred papers regarding the reliability applications with those types of censoring can be seen in Ghitany \& Al-Awadhi (2002), Goodman et. al. (2006), Joarder et. al. (2011), Iliopoulos  \& Balakrishnan (2011), Arbuckle et. al. (2014).  

The main objective of this paper is to estimate the parameters of the Weight Lindley distribution using the maximum likelihood estimation and considering data with different types of censoring, such as, type II, type I and random censoring mechanism. The originality of this study comes from the fact that, for the Weight Lindley distribution, there has been no previous work considering data with censoring mechanisms.

The paper is organized as follows. In Section 2, we review some properties of the Weight Lindley distribution. In Section 3, we present the maximum likelihood method and its properties. In Section 4, we carry out inference for this model considering different censoring mechanism. In Section 5 a simulation study is presented. In Section 6 the methodology is illustrated in a real data set. Some final comments are presented in Section 7.

\section{Weight Lindley Distribution }
\vspace{0.3cm}

Let $T$ be a random variable representing a lifetime data, with Weighted Lindley distribution and denoted by $\f{WL}(\lambda,\phi)$, the probability density function (p.d.f) is given by
\begin{equation}\label{fdpwl} 
f(t|\lambda,\phi)=\frac{\lambda^{\phi+1}}{(\lambda+\phi)\Gamma(\phi)}t^{\phi-1}(1+t)e^{-\lambda t},
 \end{equation}
for all $x>0$ , $\phi>0$ and $\lambda>0$ and $\Gamma(\phi)=\int_{0}^{\infty}{e^{-x}x^{\phi-1}dx}$ is known as gamma function. The WL (\ref{fdpwl}) distribution can be expressed as a two-component mixture
\begin{equation*}\label{fdpwl2} 
f(t|\lambda,\phi)= pf_1(t|\lambda,\phi)+(1-p)f_2(t|\lambda,\phi),
 \end{equation*}
where  $p=\lambda/(\lambda+\phi)$ and $f_j(t|\lambda,\phi)$ has p.d.f $\f{Gamma}(\phi+j-1,\lambda)$ distribution, for $j=1,2.$

The mean and variance of the WL distribution can be easily computed by
\begin{equation}\label{meanwl} 
\mu=\frac{\phi(\lambda+\phi+1)}{\lambda(\lambda+\phi)}, \sigma^2=\frac{(\phi+1)(\lambda+\phi)^2-\lambda^2}{\lambda^2(\lambda+\phi)^2}.
\end{equation}

The survival function of $T\sim WL(\theta,c)$ with the probability of an observation does not fail until a specified time $t$ is
\begin{equation}\label{fswl}
S(t|\lambda,\phi) = \frac{(\lambda+\phi)\Gamma(\phi,\lambda t)+\left(\lambda t\right)^{\phi}e^{-\lambda t}}{(\lambda+\phi)\Gamma(\phi)},
\end{equation}
where $\Gamma(x,y)=\int_{0}^{x}{w^{y-1}e^{-x}dw}$ is called upper incomplete gamma.

The hazard function quantify the instantaneous risk of failure at a given time  $t$.  The hazard function of $T$ is given by
\begin{equation}\label{fhwl} 
h(t|\lambda,\phi)=\frac{f(t|\lambda,\phi)}{S(t|\lambda,\phi)}=\frac{\lambda^{\phi+1}t^{\phi-1}(1+t)e^{-\lambda t}}{(\lambda+\phi)\Gamma(\phi,\lambda t)+\left(\lambda t\right)^{\phi}e^{-\lambda t}}. 
\end{equation}

\begin{figure}[!htb]
\centering
\includegraphics[scale=0.5]{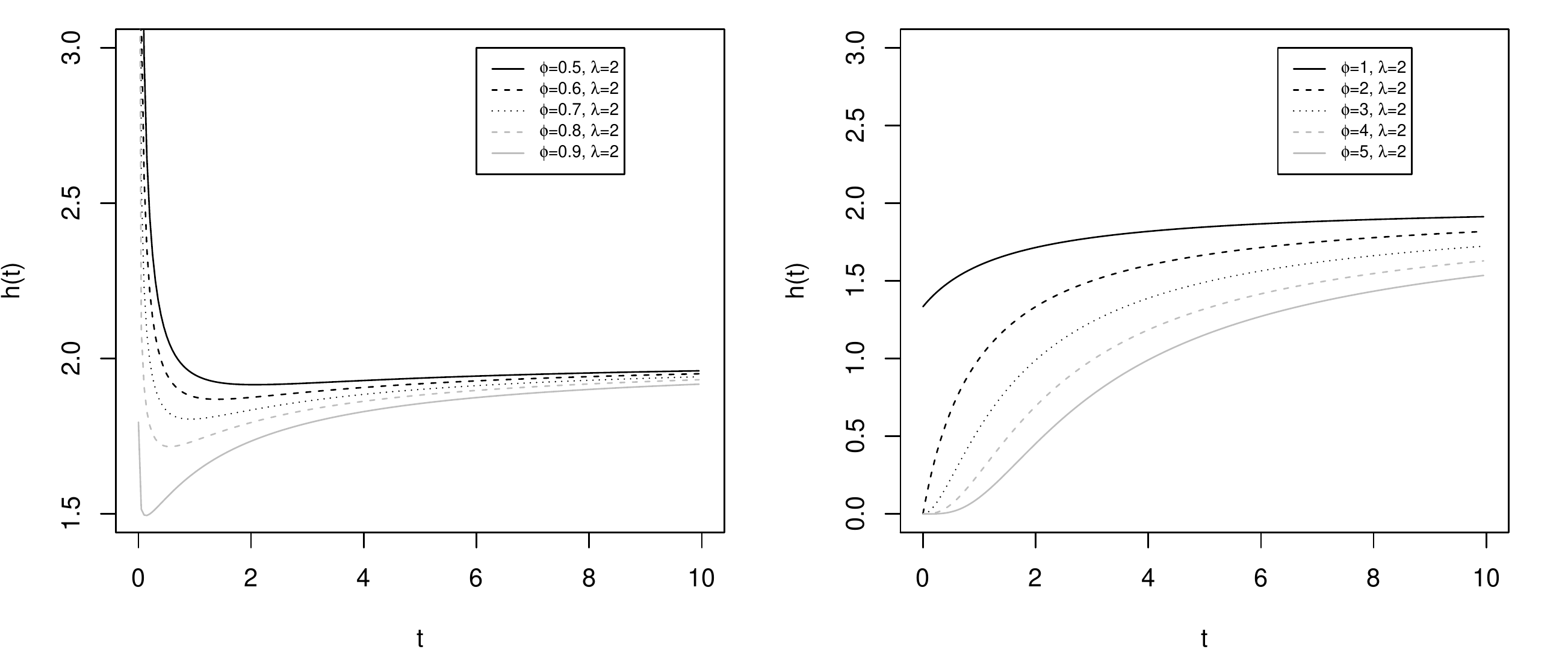}
\caption{Hazard function shapes for WL distribution and considering different values of $\phi$ and $\lambda$}\label{friswl}
\end{figure}

Figure \ref{friswl} gives examples from the shapes of the hazard function for different values of $\phi$ and $\lambda$.

\section{Maximum Likelihood Estimation }
\vspace{0.3cm}

Using the classical approach, the maximum likelihood estimators was chosen due to its asymptotic properties. Maximum likelihood estimators are obtained from maximizing the likelihood function (see, Casella e Berger, 2002). The likelihood function of $\boldsymbol{\theta}=(\theta_1,\ldots,\theta_k )$ given $t$, is
\begin{equation}\label{eqvero1} L(\boldsymbol{\theta,t})=\prod_{i=1}^n f(t_i|\boldsymbol{\theta}) . \end{equation}

For a model with $k$ parameters, if the likelihood function is differentiable at $\theta_i$, the likelihood equations are obtained by solving the equation system
\begin{equation}\label{derver}  \frac{\partial}{\partial \theta_i}\log(L(\boldsymbol{\theta,t}))=0 , i=1,2,\ldots,k.   \end{equation}
The solutions of (4) provide the maximum likelihood estimators. In many cases, numerical methods such as Newton-Rapshon are required to find the solution of the nonlinear system.

The maximum likelihood estimators of $\boldsymbol{\theta}$ are biased for small sample sizes. For large samples they are not biased and asymptotically efficient. Such estimators, under some regularity conditions, have an asymptotically normal joint distribution given by,
\begin{equation} (\boldsymbol{\hat{\theta}}) \sim N_k[(\boldsymbol{\theta}),I^{-1}(\boldsymbol{\theta})] \mbox{ para } n \to \infty , \end{equation}
where $I(\boldsymbol{\theta})$ is the Fisher information matrix, $k\times k$ and $I_{ij}(\boldsymbol{\theta})$, is the Fisher information of $\boldsymbol{\theta}$ in $i$ and $j$ given by,
\begin{equation}\label{fisherinf}
I_{ij}(\boldsymbol{\theta})=E\left[\left(\frac{\partial}{\partial \theta_i \partial \theta_j}\log(f(\boldsymbol{x|\theta}))\right)^2\right],\ i,j=1,2,\ldots,k.
\end{equation}

In the presence of censored observations, usually, its not possible to compute the Fisher information matrix, an alternative is consider the observed information matrix, where the terms is given by
\begin{equation}\label{fisherinf}
H_{ij}(\boldsymbol{\theta})=\left(\frac{\partial}{\partial \theta_i \partial \theta_j}\log(f(\boldsymbol{x|\theta}))\right)^2,\ i,j=1,2,\ldots,k.
\end{equation}

For large samples, approximated confidence intervals can be constructed for the individuals parameters $\theta_i$, with confidence coefficient $100(1-\gamma)\%$, through marginal distributions given by
\begin{equation} (\hat{\theta_i}) \sim N[(\theta_i),H^{-1}_{ii}(\boldsymbol{\theta})] \mbox{ para } n \to \infty . \end{equation}

\section{Censoring and Parameter Estimation}
\vspace{0.3cm}

In this section, we provide the maximum likelihood estimator for the two parameters of the Weight Lindley distribution considering type II, type I and random censored data. Other types of censoring such as progressive type II censoring (Balakrishnan \& Aggarwala, 2000) and Hybrid censoring mechanism (Balakrishnan \& Kundu, 2013) can also be obtained to WL distribution.

\subsection{Type II Censoring}
\vspace{0.3cm}

Usually in industrial experiments, the study of some electronic components are finished after a fixed number of failures $r$, in this case $n-r$ components will be censored. This mechanism of censoring is call type II,  see Casela \& Berger (2001) for more details, and its likelihood function is given by
\begin{equation}
L(\lambda,\phi|\boldsymbol{t})=\frac{n!}{(n-r)!}\prod_{i=1}^{r}f(t_i|\lambda,\phi)S(t_{(r)}|\lambda,\phi)^{n-r},
\end{equation}
where $t_{(r)}$ is the order statistic.

Let $T_1,\cdots,T_n$ be a random sample of WL distribution, that is, $T\sim WL(\lambda,\phi)$. The likelihood function is given by, 
\begin{equation}\label{verost2}
\begin{aligned}
L(\lambda,\phi|\boldsymbol{t})= & \frac{n!}{(n-r)!}\frac{\lambda^{r(\phi+1)}\left((\lambda+\phi)\Gamma(\phi,\lambda t_{(r)})+\left(\lambda t_{(r)}\right)^{\phi}e^{-\lambda t_{(r)}}\right)^{n-r}}{(\lambda+\phi)^{n}\Gamma(\phi)^n} \times \\ &\times \prod_{i=1}^{r}t_{(i)}^{\phi-1}(1+t_{(i)})e^{-\lambda t_{(i)}} .
\end{aligned}
\end{equation}

The logarithm of the likelihood function (\ref{verost2}) is given by, 
\begin{equation}\label{logverct2}
\begin{aligned}
l(\lambda,\phi|\boldsymbol{t})= & \sum_{i=1}^{r}\log(1+t_{(i)})+(\phi-1)\sum_{i=1}^{r}\log(t_{(i)})-\lambda\sum_{i=1}^{r}t_{(i)} - n\log\left(\Gamma(\phi)\right) \\ & +\log(n!) - \log\left((n-r)!\right)+r(\phi+1)\log(\lambda)- n\log(\phi+\lambda) \\ & +(n-r)\log\left((\lambda+\phi)\Gamma(\phi,\lambda t_{(r)})+{(\lambda t_{(r)})}^\phi e^{-\lambda t_{(r)}}\right) .
\end{aligned}
\end{equation}

From ${\partial}l(\lambda, c|\boldsymbol{t})/{\partial \lambda}=0$ and ${\partial}l(\lambda, c|\boldsymbol{t})/{\partial c}=0$, we get the likelihood equations,
{\small
\begin{equation}\label{verowl21} 
\frac{n}{\lambda+\phi}-\frac{r(\phi+1)}{\lambda}+\sum_{i=1}^{r} t_{(i)} = \frac{(n-r)\left(\Gamma(\phi,\lambda t_{(r)})e^{-\lambda t_{(r)}}-(t_{(r)}+1)\left(\lambda t_{(r)}\right)^{\phi}\right)}{\left((\lambda+\phi)\Gamma(\phi,\lambda t_{(r)})e^{-\lambda t_{(r)}}\right)+\left(\lambda t_{(r)}\right)^{\phi}},
\end{equation}
}
{\small
\begin{equation*}
\frac{(n-r)\left(\Gamma(\phi,\lambda t_{(r)})+(\lambda+\phi)\Psi(\phi,\lambda t_{(r)}) +\left(\lambda t_{(r)}\right)^{\phi}\log(\lambda t_{(r)})e^{-\lambda t_{(r)}}\right)}{\left((\lambda+\phi)\Gamma(\phi,\lambda t_{(r)})\right)+\left(\lambda t_{(r)}\right)^{\phi}e^{-\lambda t_{(r)}}} = -r\log(\lambda) 
\end{equation*}
}
{\small
\begin{equation}\label{verowl22}
+\frac{n}{\lambda+\phi} +n\psi(\phi) -\sum_{i=1}^{r}\log(t_{(i)}),
\end{equation}
}
where $\Psi(k,x)=\int_x^{\infty}w^{k-1}\log(w)e^{-w}dw=\log(x)\Gamma(k,x)+xT(3,k,x)$, and $T(m,s,x) = G_{m-1,\,m}^{\,m,\,0} \!\left( \left. \begin{matrix} 0, 0, \dots, 0 \\ s-1, -1, \dots, -1 \end{matrix} \; \right| \, x \right)$ is known as Meijer G-function. 

The solutions provide the maximum likelihood estimators of $\phi$ and $\lambda$. Numerical methods such as Newton-Rapshon are required to find the solution of the non-linear system.

\subsection{Type I Censoring}
\vspace{0.3cm}

In the presence of type I censored data, a fixed time $t_c$ is predetermined at the end of the experiment. Consider $n$ patients in a treatment and suppose that $d<n$ died until the time $t_c$, then $n-d$ are alive and will be censored. The likelihood function for this case is given by
\begin{equation}\label{eqveroc2} L(\boldsymbol{\theta,t})=\prod_{i=1}^n f(t_i|\boldsymbol{\theta})^{\delta_i}S(t_c|\boldsymbol{\theta})^{n-d} , \end{equation}
where  $d=\sum_{i}^{n}\delta_i$ is a random variable and $\delta_i=I(t_i\leq t_c)$ is an indicator function. 

Let $T_1,\cdots,T_n$ be a random sample of WL distribution, that is, $T\sim WL(\lambda,\phi)$. The likelihood function is given by, 
\begin{equation}\label{eqverodwc1}
\begin{aligned}
L(\lambda, \phi|\boldsymbol{t})=&\frac{\lambda^{d(\phi+1)}\left((\lambda+\phi)\Gamma(\phi,\lambda t_{c})+\left(\lambda t_{c}\right)^{\phi}e^{-\lambda t_{c}}\right)^{n-d}}{(\lambda+\phi)^n\Gamma(\phi)^n}\times \\ &\times\prod_{i=1}^n\left(t_{i}^{\phi-1}(1+t_{i})e^{-\lambda t_{i}}\right)^{\delta_i} .
\end{aligned}
\end{equation}

The logarithm of the likelihood function (\ref{eqverodwc1}) is given by, 
\begin{equation}\label{logverct1}
\begin{aligned}
l(\lambda,\phi|\boldsymbol{t})= &  \sum_{i=1}^{n}\delta_i\log(1+t_i)+(\phi-1)\sum_{i=1}^{n}\delta_i\log(t_i)-\lambda\sum_{i=1}^{n}\delta_it_i \\ & +(n-d)\log\left((\lambda+\phi)\Gamma(\phi,\lambda t_{c})+{(\lambda t_{c})}^\phi e^{-\lambda t_{c}}\right) \\ & +d(\phi+1)\log(\lambda)- n\log(\phi+\lambda) - n\log\left(\Gamma(\phi)\right) .
\end{aligned}
\end{equation}

From ${\partial}l(\lambda, c|\boldsymbol{t})/{\partial \lambda}=0$ and ${\partial}l(\lambda, c|\boldsymbol{t})/{\partial c}=0$, 
we get the likelihood equations,
\begin{equation}\label{verowl21} 
\frac{n}{\lambda+\phi}-\frac{d(\phi+1)}{\lambda}+\sum_{i=1}^{n}\delta_i t_i = \frac{(n-d)\left(\Gamma(\phi,\lambda t_{c})e^{-\lambda t_{c}}-(t_c+1)\left(\lambda t_{c}\right)^{\phi}\right)}{\left((\lambda+\phi)\Gamma(\phi,\lambda t_{c})e^{-\lambda t_{c}}\right)+\left(\lambda t_{c}\right)^{\phi}} ,
\end{equation}
\begin{equation*}
\frac{(n-d)\left(\Gamma(\phi,\lambda t_{c})+(\lambda+\phi)\Psi(\phi,\lambda t_{c}) +\left(\lambda t_{c}\right)^{\phi}\log(\lambda t_c)e^{-\lambda t_{c}}\right)}{\left((\lambda+\phi)\Gamma(\phi,\lambda t_{c})\right)+\left(\lambda t_{c}\right)^{\phi}e^{-\lambda t_{c}}} = -d\log(\lambda) 
\end{equation*}
\begin{equation}\label{verowl22}
+\frac{n}{\lambda+\phi} +n\psi(\phi) -\sum_{i=1}^{n}\delta_i\log(t_i).
\end{equation}

The solutions provide the maximum likelihood estimators of $\phi$ and $\lambda$. 

\subsection{Random Censoring}
\vspace{0.3cm}

In medical survival analysis and industrial life testing, random censoring schemes has been receive special attention. Suppose that the $ith$  component could experiment censoring in time $C_i$, then the data set is $(t_i,\delta_i)$, were $t_i=\min(T_i,C_i)$ and $\delta_i=I(T_i\leq C_i)$. This type of censoring have as special case type I and II censoring mechanism. The likelihood function for this case is given by
\begin{equation}\label{eqveroc2} L(\boldsymbol{\theta,t})=\prod_{i=1}^n f(t_i|\boldsymbol{\theta})^{\delta_i}S(t_i|\boldsymbol{\theta})^{1-\delta_i} .\end{equation}

Let $T_1,\cdots,T_n$ be a random sample of WL distribution, that is, $T\sim WL(\lambda,\phi)$. The likelihood function considering data with random censoring is given by, 
\begin{equation}\label{eqverodwca}
\begin{aligned} 
L(\lambda, \phi|\boldsymbol{t})=&\frac{\lambda^{d(\phi+1)}}{(\lambda+\phi)^n\Gamma(\phi)^n}\prod_{i=1}^n\left((\lambda+\phi)\Gamma(\phi,\lambda t_{i})+\left(\lambda t_{i}\right)^{\phi}e^{-\lambda t_{i}}\right)^{1-\delta_i}\times \\ & \times\left(t_{i}^{\phi-1}(1+t_{i})e^{-\lambda t_{i}}\right)^{\delta_i} .
\end{aligned}
\end{equation}

The logarithm of the likelihood function (\ref{eqverodwca}) is given by, 
\begin{equation}\label{logverctr}
\begin{aligned}
l(\lambda,\phi|\boldsymbol{t})= & \sum_{i=1}^{n}\delta_i\log(1+t_i)+(\phi-1)\sum_{i=1}^{n}\delta_i\log(t_i)-\lambda\sum_{i=1}^{n}\delta_it_i  \\ & +\sum_{i=1}^{n}(1-\delta_i)\log\left((\lambda+\phi)\Gamma(\phi,\lambda t_{i})+{(\lambda t_{i})}^\phi e^{-\lambda t_{i}}\right) \\ & +d(\phi+1)\log(\lambda)- n\log(\phi+\lambda) - n\log\left(\Gamma(\phi)\right) .
\end{aligned}
\end{equation}

From ${\partial}l(\lambda, c|\boldsymbol{t})/{\partial \lambda}=0$ and ${\partial}l(\lambda, c|\boldsymbol{t})/{\partial c}=0$,  we get the likelihood equations,
\begin{equation}\label{verowl21} 
\frac{n}{\lambda+\phi}-\frac{d(\phi+1)}{\lambda}+\sum_{i=1}^{n}\delta_i t_i = \sum_{i=1}^{n}\frac{(1-\delta_i)\left(\Gamma(\phi,\lambda t_{i})e^{-\lambda t_{i}}-(t_i+1)\left(\lambda t_{i}\right)^{\phi}\right)}{\left((\lambda+\phi)\Gamma(\phi,\lambda t_{i})e^{-\lambda t_{i}}\right)+\left(\lambda t_{i}\right)^{\phi}} ,
\end{equation}
\begin{equation*}
\sum_{i=1}^{n}\frac{(1-\delta_i)\left(\Gamma(\phi,\lambda t_{i})+(\lambda+\phi)\Psi(\phi,\lambda t_{i}) +\left(\lambda t_{i}\right)^{\phi}\log(\lambda t_i)e^{-\lambda t_{i}}\right)}{\left((\lambda+\phi)\Gamma(\phi,\lambda t_{i})\right)+\left(\lambda t_{i}\right)^{\phi}e^{-\lambda t_{i}}} = -d\log(\lambda) 
\end{equation*}
\begin{equation}\label{verowl22}
+\frac{n}{\lambda+\phi} +n\psi(\phi) -\sum_{i=1}^{n}\delta_i\log(t_i)
\end{equation}

The solutions provide the maximum likelihood estimators of $\phi$ and $\lambda$. 

\section{Simulation Study}
\vspace{0.3cm}

In this section we develop a simulation study via Monte Carlo methods. The main goal of these simulations is to study the efficiency of the proposed method. The following procedure was adopted:
\begin{enumerate}
\item Set the sample size $n$ and the parameter values $\boldsymbol{\theta}$;
\item Generate values of the $\f{WL}(\phi,\lambda)$ with size $n$;
\item Using the values obtained in step 2, calculate the MLE $\hat{\phi}$ e $\hat{\lambda}$;
\item Repeat the steps $2$ and $3$ $N$ times;
\item Using $\boldsymbol{\hat\theta}$ and $\boldsymbol{\theta}$, compute the mean relative estimates (MRE) $\sum_{i=1}^{N}\frac{\hat\theta_i/\theta_i}{N}$, the mean square errors (MSE) $\sum_{i=1}^{N}\frac{(\hat\theta_i-\theta_i)^2}{N}$, the bias $\sum_{i=1}^{N}\frac{\hat\theta_i}{N}-\theta_i$ and $95\%$ coverage probability.
\end{enumerate}

It is expected that for this approach the MRE's are closer to one with smaller MSE. The $95\%$ coverage probability was computed for the confidence intervals. For a large number of experiments, using a confidence level of $95\%$, the frequencies of intervals that covered the true values of $\boldsymbol{\theta}$ should be close $95\%$. The type II censored data were drawn setting the completed data $r$ and  $n-r$ were censored. To generate type I and random censored data, we utilize the same methods used by Goodman et. al. (2006) and Bayoud (2012), using these approaches it is expected that the proportions of censoring $E[p]$ are approximately $0.2$ and $0.4$.

The results were computed using the software R (R Core Development Team). The seed used to generate the random values was 2014. The chosen values to perform this procedure were $\boldsymbol\theta=((0.5,2),(3,2))$, $N=100,000$ and $n=(5,10,25,50,100)$. The values of $\boldsymbol\theta$ were selected to allow the increasing and bathtub shape in the hazard function. The maximum likelihood estimates were computed using the log-likelihood functions (\ref{logverct2}), (\ref{logverct1}) and (\ref{logverctr}) and the package maxLik available in R to maximize such functions. The coverage probabilities were also calculated using the numeric observed information matrix obtained from the maxLik package results.

Tables 1-6 shows the MRE's, MSE's, Bias and the coverage probability C with a confidence level equals to $95\%$ from the estimates obtained using MLE for $N$ simulated samples, considering different values of $n$, $20\%$ and $40\%$ of censored data.

\begin{table}[ht]
\centering
\caption{MRE, MSE, Bias and C estimates for $N$ samples of sizes $n=(5,10,25,50,100)$, with $20\%$ and $40\%$ of type II censored data.}
{\footnotesize
\begin{tabular}{ c c | r r r r |r r r r }
  \hline
  \multicolumn{2}{c|}{ } & \multicolumn{1}{c}{MRE} & \multicolumn{1}{c}{MSE} & \multicolumn{1}{c}{Bias} & \multicolumn{1}{c|}{$C_{95\%}$} &  \multicolumn{1}{c}{MRE} & \multicolumn{1}{c}{MSE} & \multicolumn{1}{c}{Bias} & \multicolumn{1}{c}{$C_{95\%}$}  \\ 
	\hline
	 \multicolumn{1}{c|}{n} & $r$ & \multicolumn{4}{c|}{$\phi=0.5$} & \multicolumn{4}{c}{$\lambda=2$} \\
	\hline
  \multicolumn{1}{c|}{$5$}   & \multicolumn{1}{c|}{$4$}     & 1.845 & 0.810 & 0.422 & 0.969 & 2.382 &30.240 & 2.764 & 0.961  \\
  \multicolumn{1}{c|}{$10$}  & \multicolumn{1}{c|}{$8$}     & 1.399 & 0.199 & 0.199 & 0.971 & 1.677 & 8.414 & 1.354 & 0.965  \\
  \multicolumn{1}{c|}{$25$}  & \multicolumn{1}{c|}{$20$}    & 1.131 & 0.033 & 0.065 & 0.963 & 1.217 & 1.243 & 0.435 & 0.962 \\
	\multicolumn{1}{c|}{$50$}  & \multicolumn{1}{c|}{$40$}    & 1.061 & 0.012 & 0.030 & 0.957 & 1.099 & 0.393 & 0.198 & 0.957 \\
  \multicolumn{1}{c|}{$100$} & \multicolumn{1}{c|}{$80$}    & 1.029 & 0.005 & 0.015 & 0.953 & 1.048 & 0.157 & 0.095 & 0.954\\
	\hline
  \multicolumn{1}{c|}{$5$}   & \multicolumn{1}{c|}{$3$}     & 1.950 & 0.990 & 0.475 & 0.968 & 2.920 &55.550 & 3.840 & 0.950 \\
  \multicolumn{1}{c|}{$10$}  & \multicolumn{1}{c|}{$6$}     & 1.501 & 0.281 & 0.250 & 0.971 & 2.081 &19.370 & 2.162 & 0.959 \\
  \multicolumn{1}{c|}{$25$}  & \multicolumn{1}{c|}{$15$}    & 1.177 & 0.049 & 0.089 & 0.968 & 1.383 & 3.220 & 0.767 & 0.962 \\
	\multicolumn{1}{c|}{$50$}  & \multicolumn{1}{c|}{$30$}    & 1.081 & 0.016 & 0.040 & 0.960 & 1.168 & 0.867 & 0.335 & 0.960 \\
  \multicolumn{1}{c|}{$100$} & \multicolumn{1}{c|}{$60$}    & 1.039 & 0.006 & 0.019 & 0.953 & 1.079 & 0.309 & 0.158 & 0.956 \\
	\hline
	\multicolumn{1}{c|}{n} & $r$ & \multicolumn{4}{c|}{$\phi=3$} & \multicolumn{4}{c}{$\lambda=2$} \\
	\hline
  \multicolumn{1}{c|}{$5$}   & \multicolumn{1}{c|}{$4$}     & 1.462 &10.280 & 1.385 & 0.950 & 1.489 & 4.874 & 0.978 & 0.955  \\
  \multicolumn{1}{c|}{$10$}  & \multicolumn{1}{c|}{$8$}     & 1.336 & 5.116 & 1.007 & 0.959 & 1.339 & 2.243 & 0.678 & 0.961  \\
  \multicolumn{1}{c|}{$25$}  & \multicolumn{1}{c|}{$20$}    & 1.171 & 1.802 & 0.514 & 0.959 & 1.169 & 0.753 & 0.338 & 0.961 \\
	\multicolumn{1}{c|}{$50$}  & \multicolumn{1}{c|}{$40$}    & 1.088 & 0.740 & 0.265 & 0.958 & 1.086 & 0.303 & 0.172 & 0.958 \\
  \multicolumn{1}{c|}{$100$} & \multicolumn{1}{c|}{$80$}    & 1.043 & 0.304 & 0.129 & 0.955 & 1.042 & 0.125 & 0.084 & 0.955 \\
	\hline
  \multicolumn{1}{c|}{$5$}   & \multicolumn{1}{c|}{$3$}     & 1.465 &12.250 & 1.395 & 0.944 & 1.554 & 7.022 & 1.108 & 0.947 \\
  \multicolumn{1}{c|}{$10$}  & \multicolumn{1}{c|}{$6$}     & 1.350 & 5.994 & 1.050 & 0.954 & 1.393 & 3.159 & 0.785 & 0.956 \\
  \multicolumn{1}{c|}{$25$}  & \multicolumn{1}{c|}{$15$}    & 1.208 & 2.342 & 0.623 & 0.960 & 1.224 & 1.158 & 0.447 & 0.961 \\
	\multicolumn{1}{c|}{$50$}  & \multicolumn{1}{c|}{$30$}    & 1.117 & 1.048 & 0.351 & 0.958 & 1.124 & 0.499 & 0.247 & 0.959 \\
  \multicolumn{1}{c|}{$100$} & \multicolumn{1}{c|}{$60$}    & 1.058 & 0.428 & 0.175 & 0.956 & 1.062 & 0.201 & 0.123 & 0.957 \\
	\hline
\end{tabular} }
\end{table}

\begin{table}[ht]
\centering
\caption{MRE, MSE, Bias, C and $E[p]$ estimates for $N$ samples of sizes $n=(5,10,25,50,100)$, with $20\%$ and $40\%$ of type I censored data.}
{\footnotesize
\begin{tabular}{ c|r r r r | r r r r | r }
  \hline
  \multicolumn{1}{c|}{ } & \multicolumn{1}{c}{MRE} & \multicolumn{1}{c}{MSE} & \multicolumn{1}{c}{Bias} & \multicolumn{1}{c|}{$C_{95\%}$} &  \multicolumn{1}{c}{MRE} & \multicolumn{1}{c}{MSE} & \multicolumn{1}{c}{Bias} & \multicolumn{1}{c|}{$C_{95\%}$} & \multicolumn{1}{c}{$E[p]$} \\ 
	\hline
	 $n$ & \multicolumn{4}{c|}{$\phi=0.5$} & \multicolumn{4}{c|}{$\lambda=2$} & \multicolumn{1}{r}{$ $ } \\
	\hline
  \multicolumn{1}{c|}{$5$}      & 1.740 & 0.844 & 0.370 & 0.950 & 2.084 &25.800 & 2.168 & 0.926 & 0.202 \\
  \multicolumn{1}{c|}{$10$}     & 1.301 & 0.159 & 0.151 & 0.961 & 1.420 & 4.533 & 0.841 & 0.946 & 0.199 \\
  \multicolumn{1}{c|}{$25$}     & 1.097 & 0.028 & 0.049 & 0.958 & 1.134 & 0.799 & 0.269 & 0.949 & 0.199 \\
	\multicolumn{1}{c|}{$50$}     & 1.046 & 0.011 & 0.023 & 0.955 & 1.063 & 0.313 & 0.127 & 0.949 & 0.199 \\
  \multicolumn{1}{c|}{$100$}    & 1.022 & 0.005 & 0.011 & 0.952 & 1.031 & 0.139 & 0.062 & 0.951 & 0.199 \\
	\hline
  \multicolumn{1}{c|}{$5$}      & 1.592 & 0.536 & 0.296 & 0.947 & 2.157 &27.270 & 2.314 & 0.909 & 0.390 \\
  \multicolumn{1}{c|}{$10$}     & 1.336 & 0.205 & 0.168 & 0.957 & 1.556 & 7.256 & 1.111 & 0.929 & 0.400 \\
  \multicolumn{1}{c|}{$25$}     & 1.115 & 0.039 & 0.058 & 0.959 & 1.190 & 1.478 & 0.380 & 0.941 & 0.401 \\
	\multicolumn{1}{c|}{$50$}     & 1.053 & 0.014 & 0.027 & 0.956 & 1.089 & 0.563 & 0.177 & 0.944 & 0.401 \\
  \multicolumn{1}{c|}{$100$}    & 1.026 & 0.006 & 0.013 & 0.952 & 1.043 & 0.247 & 0.087 & 0.948 & 0.401 \\
	\hline
	$n$ & \multicolumn{4}{c|}{$\phi=2$} & \multicolumn{4}{c|}{$\lambda=3$} & \multicolumn{1}{r}{$ $ } \\
	\hline
  \multicolumn{1}{c|}{$5$}      & 1.645 &25.380 & 1.935 & 0.903 & 1.658 &11.720 & 1.315 & 0.902 & 0.209 \\
  \multicolumn{1}{c|}{$10$}     & 1.307 & 5.907 & 0.922 & 0.941 & 1.296 & 2.455 & 0.591 & 0.941 & 0.205 \\
  \multicolumn{1}{c|}{$25$}     & 1.135 & 1.593 & 0.406 & 0.954 & 1.126 & 0.623 & 0.251 & 0.954 & 0.201 \\
	\multicolumn{1}{c|}{$50$}     & 1.068 & 0.663 & 0.205 & 0.954 & 1.063 & 0.261 & 0.127 & 0.954 & 0.201 \\
  \multicolumn{1}{c|}{$100$}    & 1.034 & 0.287 & 0.101 & 0.953 & 1.031 & 0.115 & 0.063 & 0.952 & 0.201 \\
	\hline
  \multicolumn{1}{c|}{$5$}      & 1.471 &17.960 & 1.413 & 0.906 & 1.526 & 9.547 & 1.053 & 0.903 & 0.404 \\
  \multicolumn{1}{c|}{$10$}     & 1.278 & 5.844 & 0.835 & 0.937 & 1.283 & 2.581 & 0.566 & 0.935 & 0.405 \\
  \multicolumn{1}{c|}{$25$}     & 1.153 & 2.076 & 0.458 & 0.951 & 1.150 & 0.908 & 0.300 & 0.949 & 0.401 \\
	\multicolumn{1}{c|}{$50$}     & 1.083 & 0.914 & 0.250 & 0.954 & 1.081 & 0.403 & 0.162 & 0.953 & 0.400 \\
  \multicolumn{1}{c|}{$100$}    & 1.042 & 0.390 & 0.125 & 0.953 & 1.041 & 0.176 & 0.082 & 0.952 & 0.400 \\
	\hline
\end{tabular} }
\end{table}

\begin{table}[ht]
\centering
\caption{MRE, MSE, Bias, C and $E[p]$ estimates for $N$ samples of sizes $n=(5,10,25,50,100)$, with $20\%$ and $40\%$ of random censored data.}
{\footnotesize
\begin{tabular}{ c|r r r r | r r r r | r }
  \hline
  \multicolumn{1}{c|}{ } & \multicolumn{1}{c}{MRE} & \multicolumn{1}{c}{MSE} & \multicolumn{1}{c}{Bias} & \multicolumn{1}{c|}{$C_{95\%}$} &  \multicolumn{1}{c}{MRE} & \multicolumn{1}{c}{MSE} & \multicolumn{1}{c}{Bias} & \multicolumn{1}{c|}{$C_{95\%}$} & \multicolumn{1}{c}{$E[p]$} \\ 
	\hline
	 $n$ & \multicolumn{4}{c|}{$\phi=0.5$} & \multicolumn{4}{c|}{$\lambda=2$} & \multicolumn{1}{r}{$ $ } \\
	\hline
  \multicolumn{1}{c|}{$5$}      & 1.887 & 1.436 & 0.443 & 0.951 & 2.282 &32.600 & 2.564 & 0.931 & 0.200 \\
  \multicolumn{1}{c|}{$10$}     & 1.330 & 0.188 & 0.165 & 0.961 & 1.490 & 5.651 & 0.980 & 0.951 & 0.201 \\
  \multicolumn{1}{c|}{$25$}     & 1.104 & 0.028 & 0.052 & 0.958 & 1.152 & 0.862 & 0.304 & 0.953 & 0.201 \\
	\multicolumn{1}{c|}{$50$}     & 1.048 & 0.011 & 0.024 & 0.955 & 1.070 & 0.317 & 0.141 & 0.952 & 0.201 \\
  \multicolumn{1}{c|}{$100$}    & 1.024 & 0.005 & 0.012 & 0.952 & 1.033 & 0.136 & 0.067 & 0.951 & 0.201 \\
	\hline
   \multicolumn{1}{c|}{$5$}     & 1.868 & 1.431 & 0.434 & 0.940 & 2.492 &45.140 & 2.984 & 0.904 & 0.409 \\
  \multicolumn{1}{c|}{$10$}     & 1.381 & 0.295 & 0.190 & 0.955 & 1.636 &10.050 & 1.271 & 0.927 & 0.418 \\
  \multicolumn{1}{c|}{$25$}     & 1.117 & 0.037 & 0.058 & 0.959 & 1.198 & 1.529 & 0.397 & 0.943 & 0.419 \\
	\multicolumn{1}{c|}{$50$}     & 1.053 & 0.013 & 0.027 & 0.956 & 1.091 & 0.554 & 0.182 & 0.947 & 0.419 \\
  \multicolumn{1}{c|}{$100$}    & 1.025 & 0.006 & 0.013 & 0.953 & 1.042 & 0.239 & 0.085 & 0.949 & 0.419 \\
	\hline
	$n$ & \multicolumn{4}{c|}{$\phi=2$} & \multicolumn{4}{c|}{$\lambda=3$} & \multicolumn{1}{r}{$ $ } \\
	\hline
  \multicolumn{1}{c|}{$5$}      & 1.960 &43.730 & 2.880 & 0.885 & 1.938 &18.630 & 1.875 & 0.890 & 0.193 \\
  \multicolumn{1}{c|}{$10$}     & 1.441 &10.330 & 1.322 & 0.936 & 1.419 & 4.027 & 0.837 & 0.937 & 0.202 \\
  \multicolumn{1}{c|}{$25$}     & 1.162 & 1.813 & 0.486 & 0.955 & 1.153 & 0.708 & 0.305 & 0.957 & 0.203 \\
	\multicolumn{1}{c|}{$50$}     & 1.077 & 0.657 & 0.231 & 0.956 & 1.073 & 0.259 & 0.146 & 0.956 & 0.203 \\
  \multicolumn{1}{c|}{$100$}    & 1.037 & 0.274 & 0.111 & 0.952 & 1.035 & 0.109 & 0.070 & 0.953 & 0.203 \\
	\hline
  \multicolumn{1}{c|}{$5$}      & 2.040 &55.450 & 3.119 & 0.861 & 2.086 &26.760 & 2.171 & 0.862 & 0.382 \\
  \multicolumn{1}{c|}{$10$}     & 1.527 &16.540 & 1.580 & 0.916 & 1.520 & 6.972 & 1.040 & 0.916 & 0.404 \\
  \multicolumn{1}{c|}{$25$}     & 1.189 & 2.501 & 0.568 & 0.951 & 1.184 & 1.053 & 0.368 & 0.952 & 0.407 \\
	\multicolumn{1}{c|}{$50$}     & 1.091 & 0.869 & 0.273 & 0.956 & 1.089 & 0.373 & 0.178 & 0.955 & 0.407 \\
  \multicolumn{1}{c|}{$100$}    & 1.043 & 0.353 & 0.130 & 0.953 & 1.042 & 0.153 & 0.084 & 0.952 & 0.407 \\
\end{tabular} }
\end{table}	
\newpage
It can be observed from the results that the Bias decreases as $n$ increases and also as expected and the values of MRE's tend to 1, allowing to get good inferences for the parameters of the Weight Lindley. That is, the estimators are asymptotically unbiased for the parameters. 
Moreover, the MSE of all estimators of the parameters tend to zero for large $n$, i.e. all estimators are consistent for the parameters.
  It is also important to point out that the coverage probabilities (C) of the two parameters approach the nominal one of $0.95$ as there is an increase of the size of $n$. 
\newpage

\vspace{0.3cm}
\section{Application}
\vspace{0.3cm}

In this section, we illustrate our proposed methodology by considering two dataset. We will consider the Weight Lindley distribution to analyze such data. For sake of comparison we obtain the results with the Weibull and Gamma distributions and nonparametric Kaplan–Meier estimator (Kaplan \& Meier, 1958). 

Firstly, to verify the behavior of the empirical hazard function it will be considered the TTT-plot (total time on test). Developed by Barlow and Campo (1975) the TTT-plot is achieve through plot of the values $[r/n,G(r/n)]$ where $G(r/n)= \left(\sum_{i=1}^{r}t_i +(n-r)t_{(r)}\right)/ {\sum_{i=1}^{n}t_i}$, $r=1,\ldots,n, i=1,\ldots,n$ and $t_{(i)}$ is the order statistics. If the curve is concave (convex), the hazard function is increasing (decreasing). When it starts convex and then concave (concave and then convex) the hazard function will have a bathtub (inverse bathtub) shape.

We also consider the AIC (Akaike Information Criteria) discrimination criterion method. Proposed by Akaike (1974) this method is based on the Kullback-Leibler Information. Let $k$ be the number of parameters to be fitted and $\hat{\boldsymbol{\theta}}$ the MLE's of $\boldsymbol{\theta}$, the AIC is computed through $AIC=-2\log(L(\hat{\boldsymbol{\theta}};\boldsymbol{t}))+2k$. Given a set of candidate models for $\boldsymbol{t}$, the preferred model is the one witch provide the minimum $AIC$ value. 

\subsection{Rats with vaginal cancer}
\vspace{0.3cm}

Presented by Pike (1966) the dataset is related to the lifetimes of $40$ rats with vaginal cancer exposed to the carcinogen DMBA. In Table 4, we reproduce the data represented by  survival times (in days) of 40 rats (+ indicates the presence of censorship).

\begin{table}[ht]
\caption{Dataset related to the lifetimes of $40$ rats with vaginal cancer exposed to the carcinogen DMBA.}
\centering 
\begin{tabular}{c c c c c c c c c c c} 
\hline 

143 & 164 & 188 & 188 & 190 & 192 & 206 & 209 & 213 & 216 \\
220 & 227 & 230 & 234 & 246 & 265 & 304+ & 216+ & 244 & 142 \\
156 & 173 & 198 & 205 & 232 & 232 & 233 & 233 & 233 & 233 \\ 
239 & 240 & 261 & 280 & 280 & 296 & 296 & 323 & 204+ & 344+ \\ [0ex] 
\hline 
\end{tabular}
\end{table}

Based on Table 4, the data clearly has random censoring mechanism, consequently the equations (\ref{verowl21}) and (\ref{verowl22}) will be used to compute the MLE's. Table 5 displays the MLE's, standard-error and  $95\%$ confidence intervals for $\phi$ and $\lambda$. Table 6 presents the results of AIC criteria, for different probability distributions.

\begin{table}[ht]
\caption{MLE, Standard-error and  $95\%$ confidence intervals for $\phi$ and $\lambda$}
\centering 
\begin{center}
  \begin{tabular}{ c | c |  c| c }
    \hline
		$\boldsymbol{\theta}$  & EMV & Erro P.& $IC_{95\%}(\theta)$ \\ \hline
    \ \ $\phi$ \ \   & 21.7545 & 1.3254 &  (19.1566; 24.3523)  \\ \hline
    \ \ $\lambda$   \ \  &  0.0978 &  0.0066 &  (0.0848; 0.1109) \\ \hline
  \end{tabular}
\end{center}
\end{table}

\begin{table}[ht]
\caption{Results of AIC criteria for different probability distributions considering the lifetimes of $40$ rats with cancer.}
\centering 
\begin{center}
  \begin{tabular}{ c | c | c | c}
    \hline
		Criteria & Weight Lindley & \ Weibull \ & \ \ Gamma \\ \hline
      AIC & \textbf{390.3428} & 394.4234 & 390.6482 \\ \hline
  \end{tabular}
\end{center}
\end{table}

\begin{figure}[!htb]
\centering
\includegraphics[scale=0.45]{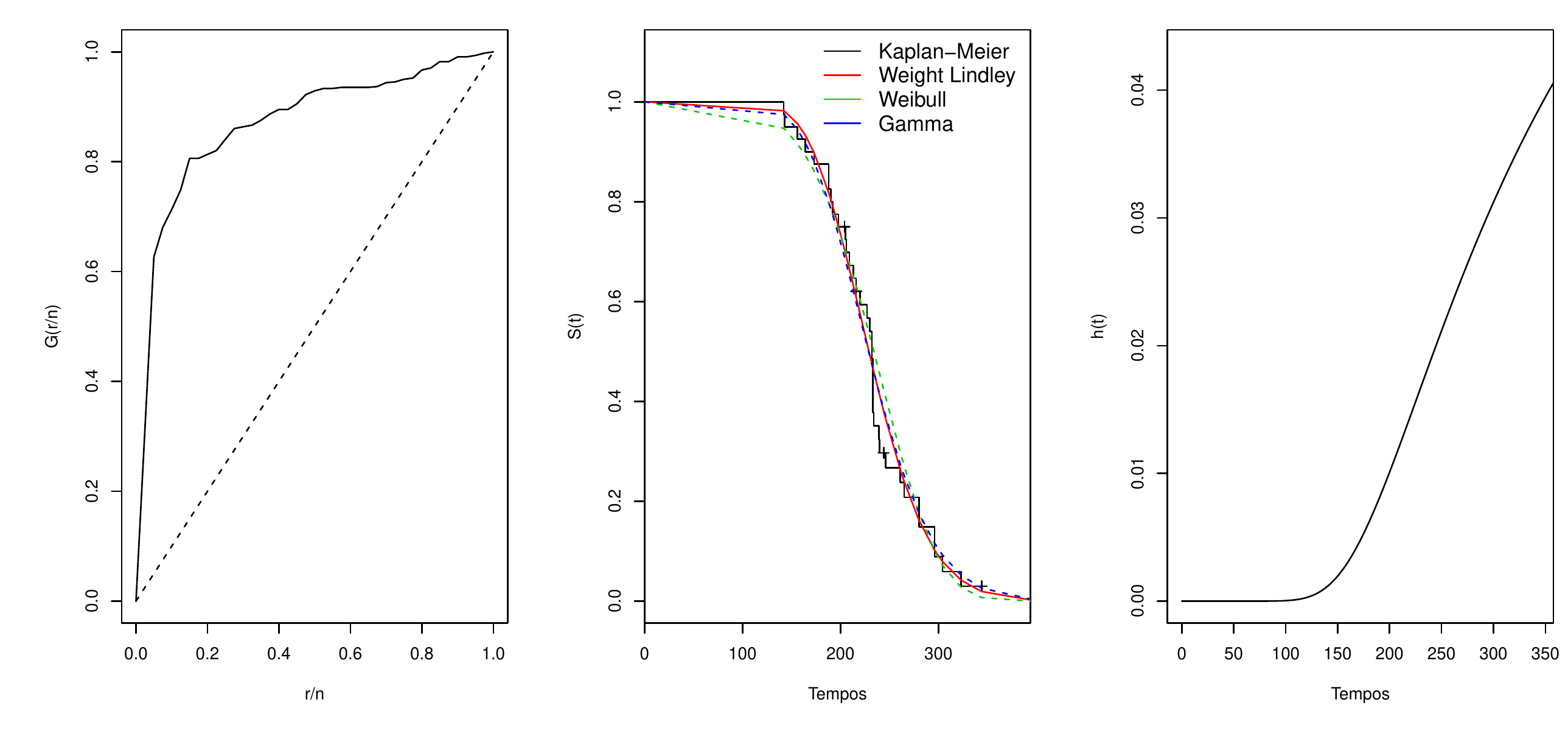}
\caption{TTT-plot, survival function adjusted by different distributions and Kaplan–Meier estimator and the hazard function adjusted by WL distribution considering the lifetimes of $40$ rats with cancer.}\label{grafico-obscajust1}
\end{figure}

In the Figure \ref{grafico-obscajust1}, we have the TTT-plot, survival function adjusted by different distributions and Kaplan–Meier estimator and the hazard function adjusted by WL distribution.

Based on the TTT-plot there is a indication that the hazard function has increasing shape. Comparing the empirical survival function with the adjusted distributions it can be observed a good fit for the Weight Lindley distribution. These result is confirmed from AIC since WL distribution has the minimum value. The hazard function adjusted by WL distribution confirms the result obtained from TTT-plot. Therefore, through the proposed methodology the data related to rats with vaginal cancer can be described by the Weight Lindley distribution.

\subsection{Lifetime of electrical devise}
\vspace{0.3cm}

Presented by Lawless (2002, p.112) the dataset is related to $60$ electrical devices. The survival times is given in cycles to failure divided by $1000$ and was firstly presented without censoring. It will be considered that the experiment was ended after we observe $r=49$ failure, therefore $n-r=11$ components will be censored. Table 7 reproduces the lifetimes from the first 49 electrical devices.

\begin{table}[ht]
\caption{Dataset related to the lifetimes of $60$ (in cycles) electrical devices.}
\centering 
\begin{tabular}{c c c c c c c c c } 
\hline 

0.014 & 0.034 & 0.059 & 0.061 & 0.069 & 0.080 & 0.123 & 0.142 & 0.165 \\
0.210 & 0.381 & 0.464 & 0.479 & 0.556 & 0.574 & 0.839 & 0.917 & 0.969 \\ 
0.991 & 1.064 & 1.088 & 1.091 & 1.174 & 1.270 & 1.275 & 1.355 & 1.397 \\ 
1.477 & 1.578 & 1.649 & 1.702 & 1.893 & 1.932 & 2.001 & 2.161 & 2.292 \\ 
2.326 & 2.337 & 2.628 & 2.785 & 2.811 & 2.886 & 2.993 & 3.122 & 3.248 \\ 
3.715 & 3.79 & 3.857 & 3.912 \\ [0ex] 
\hline 
\end{tabular}
\end{table}

The experiment was ended after a predetermined number of failures $r$, therefore the data has type II censoring mechanism and the equations (\ref{verowl21}) and (\ref{verowl22}) will be used to compute the MLE's. Table 8 displays the MLE's, standard-error and  $95\%$ confidence intervals for $\phi$ and $\lambda$. Table 9 presents the results of AIC criteria, for different probability distributions, considering the electrical devices data.

\begin{table}[ht]
\caption{MLE, Standard-error and  $95\%$ confidence intervals for $\phi$ and $\lambda$ considering the electrical devices data}
\centering 
\begin{center}
  \begin{tabular}{ c | c |  c| c }
    \hline
		$\boldsymbol{\theta}$  & EMV & Erro P.& $IC_{95\%}(\theta)$ \\ \hline
    \ \ $\phi$ \ \   & 0.6764 & 0.1341 &  (0.4137; 0.9392)  \\ \hline
    \ \ $\lambda$   \ \  &  0.5260 &  0.0954 &  (0.3391; 0.7129) \\ \hline
  \end{tabular}
\end{center}
\end{table}

\begin{table}[ht]
\caption{Results of AIC criteria for different probability distributions considering the electrical devices data.}
\centering 
\begin{center}
  \begin{tabular}{ c | c | c | c}
    \hline
		Criteria & Weight Lindley & \ Weibull \ & \ \ Gamma \\ \hline
      AIC & \textbf{185.1739} & 186.5965 & 186.218 \\ \hline
  \end{tabular}
\end{center}
\end{table}

\begin{figure}[!htb]
\centering
\includegraphics[scale=0.45]{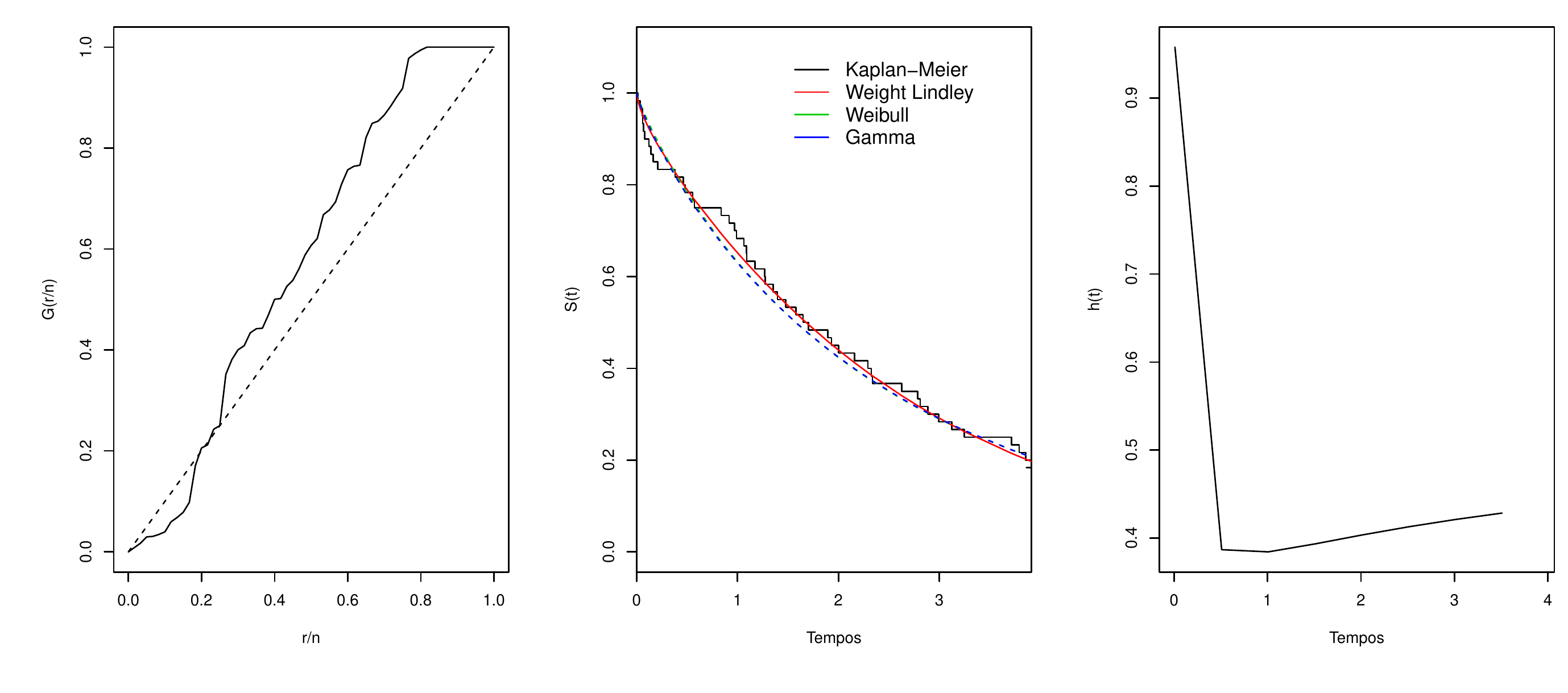}
\caption{TTT-plot, survival function and the hazard function adjusted by Weight Lindley distribution considering the electrical devices data}\label{grafico-obscajust2}
\end{figure}

In the Figure \ref{grafico-obscajust2}, we have the TTT-plot, survival function adjusted by different distributions and Kaplan–Meier estimator and the hazard function adjusted by WL distribution.

Similar to first dataset, comparing the empirical survival function with the adjusted distributions and through the AIC results it can be observed a good fit for the Weight Lindley distribution. Based on the TTT-plot there is a indication that the hazard function has bathtub shape. The hazard function adjusted by Weight Lindley Distribution confirm those results. Therefore, through the proposed methodology the data considering the electrical devices can be described by the Weight Lindley distribution.

\section{Final Comments}
\vspace{0.3cm}

In this paper, we derived the maximum likelihood equations for the parameters of the Weight Lindley distribution considering different types of censoring, such as, type I, type II and random censoring mechanism.

Based on simulation studies and on real applications, we demonstrated that using  the proposed methodology it was possible to obtain good estimates of the parameters of Weight Lindley distribution. These results are of great practical interest since this will enable us for the use of the Weight Lindley distribution in various application issues.

There are a large number of possible extensions of the current work. 
The presence of covariates, as well as  of long-term survivals, is very common in practice. Our approach should be investigate in both contexts. A possible approach is to consider the regression schemes adopted by Achcar \& Louzada-Neto (1992) and Perdona \& Louzada-Neto (2011), respectively.

\section*{Acknowledgements}

The research was partially supported by CNPq, FAPESP and CAPES of Brazil.

\bibliographystyle{abbrv}

\end{document}